\documentclass[aps,prb,amsmath,amssymb,showpacs,twocolumn,floatfix]{revtex4-1}
\usepackage{epsfig, color}
\usepackage{amsfonts, relsize}
\usepackage{graphics}
\usepackage{graphicx}
\usepackage{hyperref}

\begin{document}

\title{Axionic superconductivity in three dimensional doped narrow gap semiconductors}

\author{Pallab Goswami}
\affiliation{ National High Magnetic Field Laboratory, Florida State University, FL 32306, USA}

\author{Bitan Roy}
\affiliation{ National High Magnetic Field Laboratory, Florida State University, FL 32306, USA} 
\affiliation{Condensed Matter Theory Center, Department of Physics, University of Maryland, College Park, MD 20742, USA}

\date{\today}

\begin{abstract}
We consider the competition between the conventional s-wave and the triplet Balian-Werthamer or the B-phase pairings in the doped three dimensional narrow gap semiconductors, such as $\mathrm{Cu}_x\mathrm{Bi}_2\mathrm{Se}_3$ and $\mathrm{Sn}_{1-x}\mathrm{In}_x\mathrm{Te}$. When the coupling constants of the two contending channels are comparable, we find a simultaneously time-reversal and parity violating $p + is$ state at low temperatures, which provides an example of dynamic axionic state of matter. In contradistinction to the time-reversal invariant, topological B-phase, the $p + is$ state possesses gapped Majorana fermions as the surface Andreev bound states, which give rise to an anomalous surface thermal Hall effect. The anomalous gravitational and electrodynamic responses of the $p+is$ state can be described by the $\theta$ vacuum structure, where $\theta \neq 0$ or $\pi$.
\end{abstract}


\maketitle

\vspace{10pt}

The notion of $\theta$ vacuum is a venerable concept of the modern quantum field theory, which has profound implications for the vacuum structure of the gauge theories \cite{Rajaraman}. Physically, the $\theta$ term for the gauge theory is a \emph{pseudoscalar magneto-electric coefficient}, and an arbitrary $\theta$ violates the time reversal ($\mathcal{T}$) and the parity ($\mathcal{P}$) symmetries. In order to preserve $\mathcal{P}$ and $\mathcal{T}$ symmetries in the strong interaction regime and also to account for the violation of these fundamental discrete symmetries in the lower energy scales, the existence of a very light pseudoscalar boson, dubbed \emph{axion} was postulated almost three decades ago \cite{Peccei, Weinberg, Wilczek1}. Thus far the axion has eluded the experimental detection.

However, the interest in the $\theta$ vacuum and its tangible experimental consequences has been revived due to the discovery of the time reversal symmetric (TRS), three dimensional $Z_2$ topological insulators \cite{AxionHughes, Review1, Review2}. It has been recognized that the topological invariant of the $Z_2$ topological insulator couples to the electromagnetic gauge field as a $\theta$ term, and as a consequence of the $\mathcal{T}$ symmetry and the non-degeneracy of the underlying ground state, the magneto-electric coefficient $\theta$ is quantized to be $\pi$ \cite{AxionHughes}. When the criterion of the non-degenerate ground state is relaxed, while maintaining the $\mathcal{T}$ symmetry, the $\theta$ can acquire fractional values, which reflect the degeneracy of the ground state on a torus \cite{Maciejko,Senthil}. The similar $\theta$ vacuum structures have also been found for the spin gauge field \cite{Ryu} and the gravitational field \cite{GravQi,GravRyu1} respectively for the TRS topological superconductors (TSC) in the classes CI and DIII \cite{TenfoldRyu}.

The Balian-Werthamer or the B-phase of superfluid $^3He$ \cite{TenfoldRyu} and the pseudoscalar pairing of the four component charged Dirac fermions \cite{FuBerg} are experimentally pertinent examples of TSC in the class DIII. Upon projection onto the low energy quasiparticles in the vicinity of the Fermi surface, the pseudoscalar pairing also maps onto the B-phase \cite{Silaev}. Following the suggestion of Ref.~\onlinecite{FuBerg}, that the DIII TSC may be realized in doped three dimensional, strongly spin-orbit coupled narrow gap semiconductors, there has been considerable experimental interest in the superconducting $\mathrm{Cu}_x\mathrm{Bi}_2\mathrm{Se}_3$ and $\mathrm{Sn}_{1-x}\mathrm{In}_x\mathrm{Te}$ \cite{ObWray1, ObKriener, ObWray2,ObSasaki,PCSasaki, PCKirzhner, PCChen, ObLevy, Novak, Tranquada}.

A natural question arises, whether there is a condensed matter realization of $\mathcal{P}$ and $\mathcal{T}$ breaking dynamic axionic state of matter \cite{Wilczek2}. Recently, the possibility of such a phase has been proposed for some magnetic insulators \cite{Essin, LiQi1,LiQi2}, which is yet to be experimentally found. In this paper, we demonstrate that a spontaneously $\mathcal{P}$ and $\mathcal{T}$ breaking \emph{axionic superconducting state} can be realized in the doped three dimensional narrow gap semiconductors. This axionic paired state has $p+is$ pairing symmetry, and emerges due to the competition between the conventional s-wave and the triplet B-phase pairings,  and lacks any analog in the superfluid $^3He$ \cite{Volovik}.  According to the Altland-Zirnbauer classification scheme the $p+is$ state is a member of the class D \cite{Altland}. Both the electromagnetic gauge field and the axion fields are massive inside this phase. As a consequence of the broken $\mathcal{T}$, the $p + is$ state possesses gapped Majorana fermions as the surface Andreev bound states (SABS), and supports an anomalous surface thermal Hall (STH) conductivity.

The low energy quasi-particle dispersion in many narrow gap semi-conductors is succinctly captured by a massive Dirac equation, which describes the Kramer's degenerate quasi-particles in the conduction and the valence bands. In the presence of strong spin-orbit coupling, electron-phonon scattering can lead to attractive interactions in both the singlet and the triplet channels \cite{FuBerg}. As these materials are weakly correlated, perhaps it is not a strong assumption that the retarded pairing interaction will generically emerge in the vicinity of the Fermi surface. Therefore, the completely filled (or empty) bands can be safely integrated out for determining the low energy pairing physics. We consider a situation, when the Fermi level lies in the conduction band, and begin with the following interacting Hamiltonian
\begin{eqnarray}
H_{qp}=\sum_{\mathbf{k},s} \xi_{k}\: c_{\mathbf{k},s}^{\ast}  c_{\mathbf{k},s}+\sum_{\mathbf{k}}V(\mathbf{k})\: \hat{n}_{\mathbf{k}} \; \hat{n}_{-\mathbf{k}}.
\end{eqnarray}
In the above equation the number operator $\hat{n}_{\mathbf{k}}=\sum_{\mathbf{q},s}c_{\mathbf{q}+\mathbf{k},s}^{\ast}c_{\mathbf{q},s}$, and $c_{\mathbf{k},s}^{\ast}$, $c_{\mathbf{k},s}$ are respectively the creation and annihilation operators of the quasi-particles with momentum $\mathbf{k}$. The index $s$ represents the Kramer's pair, and $\xi_{\mathbf{k}}=\sqrt{v^2 k^2 +\Delta^2_g}-\mu$ describes the quasi-particle energy with respect to the Fermi level $\mu$. The band gap is denoted by $\Delta_g$ and the band parameter $v$ has the dimension of velocity. When $|\mu-\Delta_g| \ll \Delta_g$, a non-relativistic approximation $\xi_{\mathbf{k}} \approx k^2/(2m)-\tilde{\mu}$ can be applied, where $m=\Delta_g$ and $\tilde{\mu}=\mu-\Delta_g$. If we choose a simplified interaction potential $V(\mathbf{k})$, which is a sum of the attractive interactions in the s-wave and the p-wave channels, the pertinent reduced BCS Hamiltonian for the mean-field description becomes $H=\frac{1}{2}\sum_{\mathbf{k}} \Psi^{\dagger}_{\mathbf{k}} \hat{H}_{\mathbf{k}}\Psi_{\mathbf{k}}$. The four component Nambu spinor $\Psi^{\dagger}_{\mathbf{k}}=\left( c_{\mathbf{k},\uparrow}^{\ast}, c_{\mathbf{k},\downarrow}^{\ast}, c_{-\mathbf{k},\downarrow}, - c_{-\mathbf{k},\uparrow} \right)$, and the operator
\begin{eqnarray} \label{reducedBCS1}
\hat{H}_{\mathbf{k}} =\left(
\begin{array}{c c}
\xi_k \; \sigma_0 &  \Delta_s \; \sigma_0 + \Delta_t \; \mathbf{d}_{\mathbf{k}}\cdot \boldsymbol \sigma  \\
\Delta_s^{\ast} \; \sigma_0 + \Delta_t \; \mathbf{d}^{\ast}_{\mathbf{k}}\cdot \boldsymbol \sigma  & -\xi_k \; \sigma_0
\end{array}
\right),
\end{eqnarray}
where $\sigma_0$ and $\boldsymbol \sigma$ respectively denote the identity and the conventional Pauli matrices operating on the Kramer's indices. We have introduced the complex s-wave pairing amplitude $\Delta_s$, and a triplet amplitude $\Delta_t$. In a weak coupling approach, the time-reversal symmetric (TRS), fully gapped B-phase is energetically most favorable in the triplet channel, and is characterized by $\mathbf{d}_{\mathbf{k}}=\mathbf{k}/k_F$, where $k_F$ is the Fermi momentum.

\begin{figure}[htb]
\includegraphics[width=8.5cm,height=5cm]{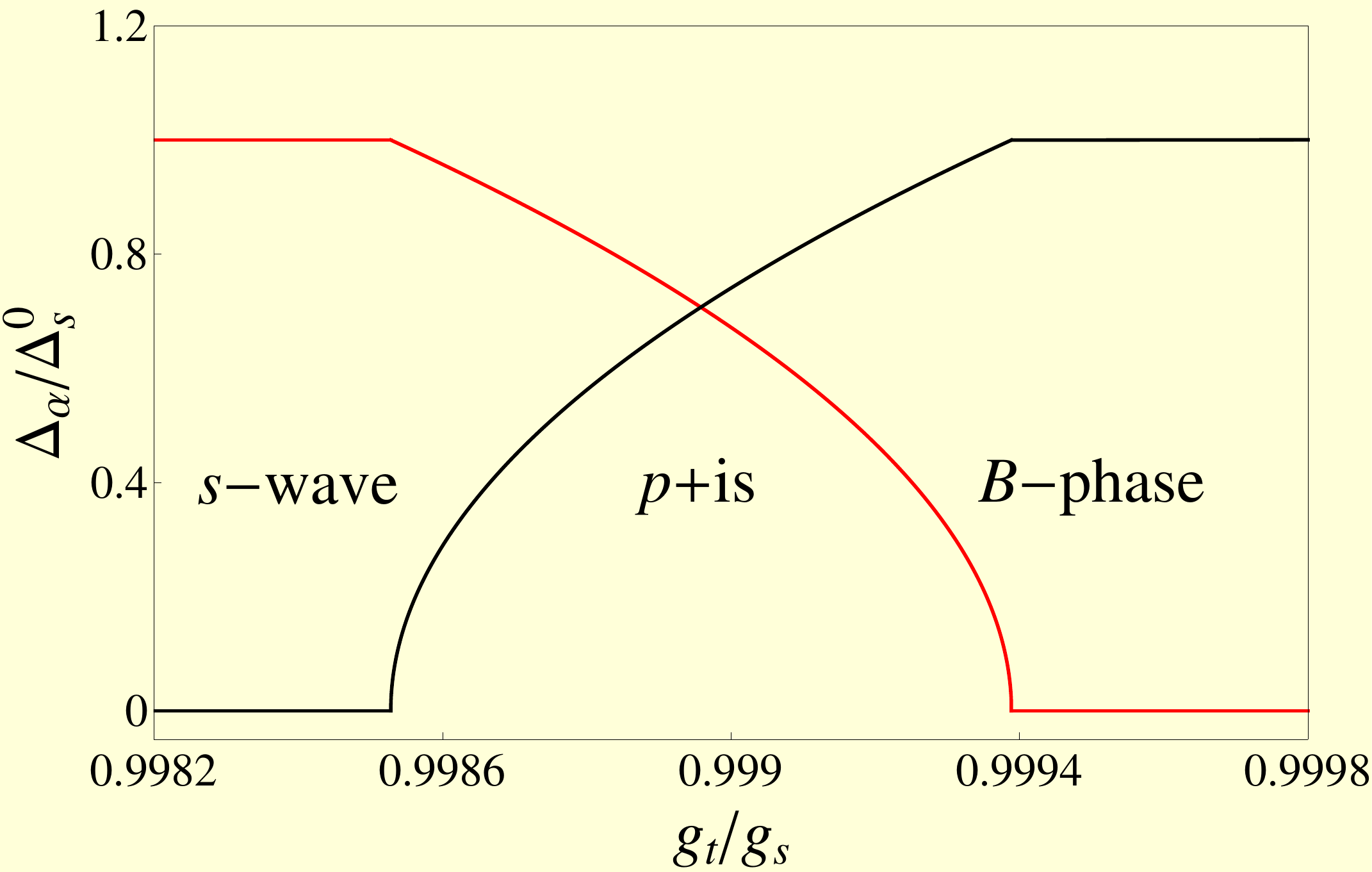}
\caption{(Color online)A cut of the zero temperature phase diagram for $g_s \rho(\mu)=1$, showing the normalized pairing amplitudes, as a function of the ratio $g_t/g_s$, where $\rho(\mu)$ is the density of states at the Fermi level. The amplitudes are normalized by $\Delta^0_s$, the s-wave amplitude, when the triplet channel is turned off. The s-wave and the B-phase amplitudes are respectively shown as the red and the black lines.}
\label{coexistence}
\end{figure}

When we focus on the competition between the s-wave and the B-phase pairings, the associated reduced BCS Hamiltonian is described by
\begin{eqnarray} \label{reducedBCS3}
\hat{H}_{\mathbf{k}} =\xi_{\mathbf{k}} \gamma_0+\frac{\Delta_t}{k_F}\; \gamma_0 \gamma_j k_j + \Re{(\Delta_s)} \gamma_5 + i \; \Im{(\Delta_s)} \gamma_0 \gamma_5,
\end{eqnarray}
where $\gamma_0=\sigma_0 \otimes \tau_3$, $\gamma_j=i \sigma_j \otimes \tau_2$, $\gamma_5=\sigma_0 \otimes \tau_1$, and $\tau_j$ are the Pauli matrices operating in the Nambu space. A pristine s-wave phase only breaks $U(1)$ gauge symmetry, and leaves all the discrete symmetries intact, and is topologically trivial. In the absence of any s-wave pairing, the Hamiltonian for the B-phase takes the form of a \emph{band inverted, massive Dirac equation in three dimensions}. Due to the presence of $\mathcal{T}$ and the broken spin rotational symmetry, this phase belongs to the class DIII, and is characterized by an integer topological invariant
\begin{equation}
\mathcal{N}=\frac{1}{2}[1+\mathrm{sgn}(m\tilde{\mu})]\; \mathrm{sgn}(\Delta_t),
\end{equation}
which for weak or BCS pairing ($\mathrm{sgn}(m\tilde{\mu})=1$) reduces to $\mathcal{N}=\mathrm{sgn}(\Delta_t)$. On the other hand, for strong pairing or BEC limit ($\mathrm{sgn}(m\tilde{\mu})=-1$), $\mathcal{N}=0$. As a consequence of the $\mathcal{T}$ symmetry, the Hamiltonian of the B-phase involves only four mutually anti-commuting Dirac $\gamma$ matrices, and the gravitational response of the B-phase is characterized by an \emph{axion angle} $\theta^0_{ax}=\pi$ \cite{GravQi,GravRyu1}. It is important to note that the B-phase pairing breaks the inversion symmetry of the normal state. However, due to the spin-orbital locking in the B-phase, there is an \emph{emergent parity} symmetry $\mathcal{P}$ defined by $\mathbf{k} \to -\mathbf{k}$ and $\Psi_{\mathbf{k}} \to \gamma_0 \Psi_{\mathbf{k}}$. When both pairings coexist, the imaginary part of the s-wave amplitude appears as a \emph{pseudo-scalar mass}, with the fifth anti-commuting $\gamma$ matrix, and enhances the gap on the Fermi surface. On the other hand, the real part of the s-wave pairing commutes with the B-phase operator, and only anti-commutes with the kinetic energy. Therefore, the real part appears as an \emph{axial chemical potential} \cite{GoswamiTewari}. In the coexisting phase, both $\Re{(\Delta_s)}$ and $\Im{(\Delta_s)}$ break the $\mathcal{P}$ symmetry, and only the \emph{pseudo-scalar mass} simultaneously breaks $\mathcal{P}$ and $\mathcal{T}$.

The emergence of the $p+is$ state can be justified in the following manner. The quasi-particle spectra
corresponding to $\hat{H}_{\mathbf{k}}$ in Eq.~\ref{reducedBCS3} are given by $E_{\mathbf{k}}=\pm E_{\alpha,\mathbf{k}}$, and
\begin{equation}\label{spectrum}
E_{\alpha,\mathbf{k}}=
\sqrt{\xi^2_k + |\Delta_s|^2+  \Delta^2_t \frac{k^2}{k^2_F} + 2 \; \alpha \; \Delta_t \frac{k}{k_F} \; \Re{(\Delta_s)}},
\end{equation}
where $\alpha = \pm 1$. Form the above dispersion relations it becomes clear that the gap is maximized, when $\Delta_s$ is purely imaginary, and leads to an axion angle $\theta^0_{ax}=\pi + \tan^{-1} \Im{(\Delta_s)}/\mu$, for the gravitational response. Due to the simultaneous violation of $\mathcal{T}$ and the spin rotational symmetries, the $p+is$ state belongs to the class D \cite{TenfoldRyu, Altland}.

The stabilization of $p+is$ phase can be further substantiated via a minimization of the free energy
\begin{eqnarray}\label{freeenergy}
f_s=\frac{|\Delta_s|^2}{2g_s}+\frac{|\Delta_t|^2}{2g_t}-2 T \sum_{\alpha,\mathbf{k}}\log \left[2\cosh \frac{E_{\alpha,\mathbf{k}}}{2 T}\right]+\sum_{\mathbf{k}} \xi_{\mathbf{k}}, \nonumber \\
\end{eqnarray}
where $g_s$, $g_t$ are respectively the coupling constants in the s-wave and the p-wave channels, and $T$ is the temperature (throughout the Letter we are using the units $e=c=\hbar=k_B=1$). We illustrate our findings through a cut of the phase diagram at $T=0$, as a function of the ratio $g_t/g_s$ in Fig. 1. For simplicity, we have chosen the same energy cut-off $\omega_D=0.1 \mu$ in both pairing channels, and we are demonstrating the results for $g_s \rho (\mu)=1$, where $\rho(\mu)$ stands for the density of states at the Fermi level. Since both the s-wave and the B phase are fully gapped states, the coexistence occurs only in a sliver of the entire phase diagram, when $g_t/g_s \sim 1$. This phase diagram suggests the presence of two stage thermal phase transitions in the vicinity of $g_t/g_s \sim 1$. As the temperature is gradually lowered, one first enters the dominant pure phase depending on the relative strength of the couplings, and the $\mathcal{T}$ breaking occurs only at a lower temperature.

When the transition temperatures of the two pairings are comparable, the coexistence can be addressed by using the phenomenological Landau-Ginzburg free energy. The condensation energy density $\Delta f=f_s-f_n$ can be written as
\begin{eqnarray}\label{Landaufree}
\Delta f&=& \sum_{\alpha=s,t}\left[\frac{c_{\alpha}}{2}|(\nabla-2i\mathbf{A})\Delta_{\alpha}|^2+r_{\alpha} |\Delta_{\alpha}|^2 + u_{\alpha} |\Delta_{\alpha}|^4\right] \nonumber \\&+& u_{st1} |\Delta_s|^2 |\Delta_t|^2 + u_{st2} |\Delta_s|^2 |\Delta_t|^2 \cos 2\theta_-  +\frac{\mathbf{B}^2}{8\pi},
\end{eqnarray}
where $f_n$ is the normal state's free energy density, and $\theta_-=(\theta_t-\theta_s)$ is the relative phase between the two complex amplitudes $\Delta_{\alpha}=|\Delta_{\alpha}|\exp (i\theta_{\alpha})$, and $\mathbf{B}=\nabla \times \mathbf{A}$ is the magnetic field strength. The constants $c_{\alpha}$ have the dimension of inverse mass and the individual superfluid stiffness can be defined as $\rho_{\alpha}=c_{\alpha} |\Delta_{\alpha}|^2$. As shown in the Supplementary Material \cite{Supplementary}, all the quartic coefficients for this problem turn out to be positive definite, and consequently the free energy is minimized for $\theta_-= \pm \pi/2$ in the coexisting phase. Deep inside the $p+is$ phase we can ignore the amplitude fluctuations, and in the absence of any singularity in the phase fields we can also shift the vector potential as $\mathbf{A} \to \mathbf{A} -\frac{\nabla \theta_+}{2e}+\frac{\rho_-}{2\rho_+}\nabla \theta_-$, where $\rho_{\pm}=\rho_s \pm \rho_t$ and $\theta_{+}=(\theta_s+\theta_t)/2$. After this shift, the explicit form of the free energy in the London limit becomes
\begin{eqnarray}
\Delta f=\frac{\rho^2_+-\rho^2_-}{8\rho_+}\left[ \left(\nabla \delta\theta_-\right)^2- \frac{2\; u_{st2}\; \rho_+}{c_s \; c_t} \cos (2 \; \delta \theta_-)\right] \nonumber \\ +
\frac{\rho_+}{2}\mathbf{A}^2 +\frac{\mathbf{B}^2}{8\pi},
\end{eqnarray}
where $\delta \theta_-$ is the deviation of $\theta_-$ from $\pm \pi/2$, and this equation shows that both the gauge field and the \emph{axionic excitations} are massive.

\begin{figure}[htb]
\includegraphics[width=8.75cm,height=7cm]{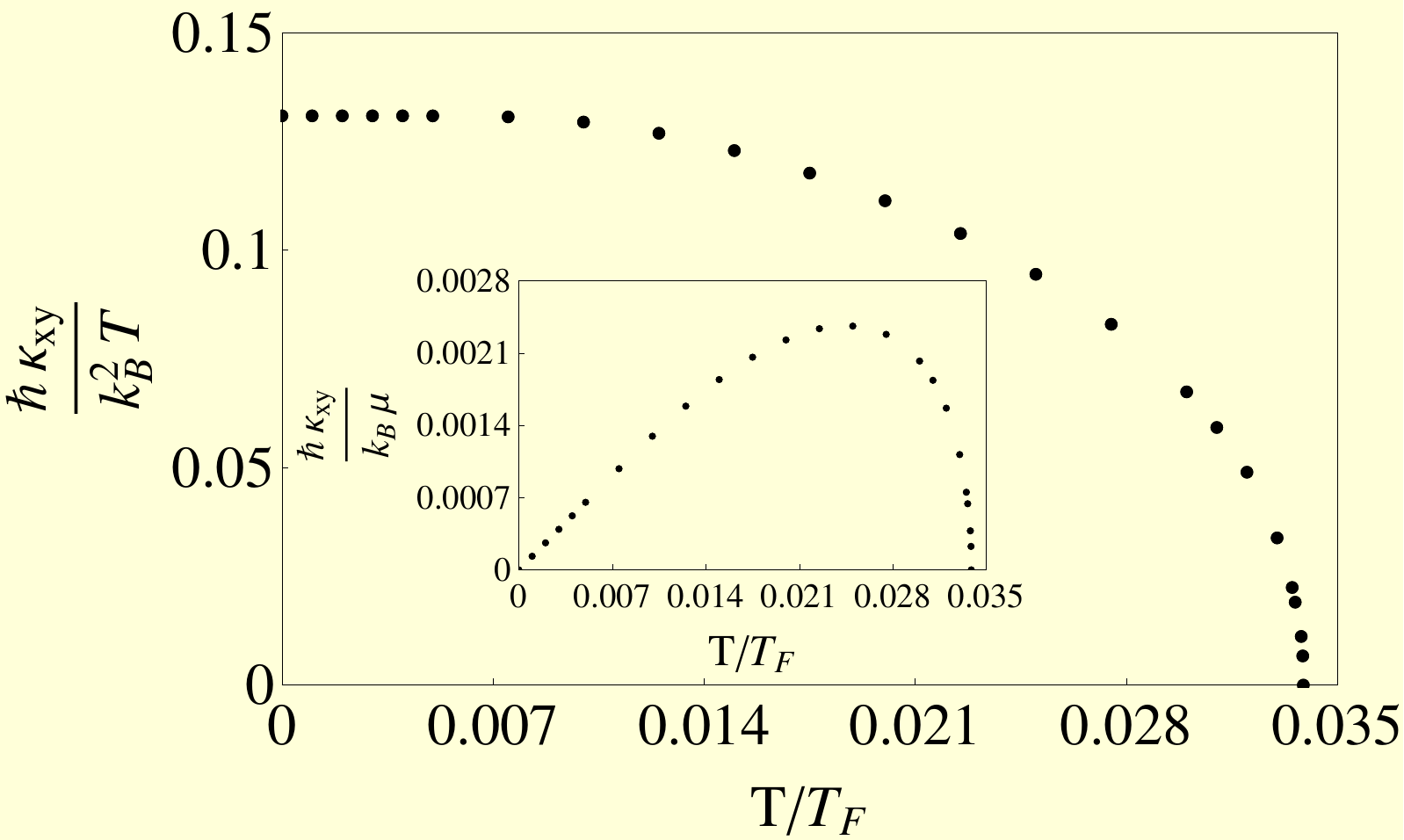}
\caption{(Color online)The dimensionless quantity $\hbar \kappa_{xy}/(k^2_B T)$ for the gapped SABS, as a function of $T/T_F$, where $\kappa_{xy}$ and $T_F=\mu/k_B$ are respectively the thermal Hall conductivity and the Fermi temperature. We have used the mean-field gap amplitudes for $g_s\rho(\mu)=1$ and $g_t/g_s=0.99895$. This quantity saturates to the universal number $\pi/24$ in the zero temperature limit. Inset: The dimensionless thermal Hall conductivity $\hbar \kappa_{xy}/(k_B \mu)$ for the gapped SABS, as a functions of $T/T_F$.}
\label{coexistence}\end{figure}

Next we focus on the physical implications of this axionic superconductor. We begin by demonstrating the existence of the \emph{gapped} SABS, which in turn lead to an anomalous STH effect. For concreteness, we assume that the semi-infinite regions with $z<0$ and $z>0$ are respectively occupied by the $p+is$ superconductor and the vacuum. Therefore, the spinor wave-function must satisfy the boundary conditions $\psi(z=0)=\psi(z\to -\infty)=0$. For simplicity, we also choose $\xi_{\mathbf{k}}=k^2/(2m)-\mu$ and set $\Re(\Delta_s)=0$ in Eq.~\ref{reducedBCS3}. Now assuming $\psi(z) \sim \exp (\lambda z)$, and subsequently setting $k_z \to i \lambda$ in Eq.~\ref{spectrum}, we obtain the following secular equation for $\lambda$
\begin{eqnarray}\label{secularlambda}
\lambda^4 + 2 \lambda^2 \left[ k^2_F- k^2_\perp - 2m^2v^2_\Delta\right] +(k^2_F- k^2_\perp)^2 \nonumber \\ + 4m^2\left(v^2_\Delta k^2_\perp +\Im (\Delta_s)^2-E^2\right)
=0,
\end{eqnarray}
where $v_\Delta=\Delta_t/k_F$, and $E$ is the energy of the SABS. The solutions of this equation are of the form $\pm \lambda_j$, with $j=1,2$. In order to satisfy $\psi(z\to -\infty)=0$, we require $\Re(\lambda_j)>0$. After imposing the condition $\psi(z=0)=0$, we obtain the constraints
 \begin{eqnarray}\label{energylambda}
4mv^2_\Delta \Delta e^{-i  \phi}\left(m \Delta e^{-i  \phi}-  k^2_\perp - \lambda_1 \lambda_2\right) + 2 E^2 \lambda_1 \lambda_2 \nonumber \\
 +v^2_\Delta(k^4_\perp  + \lambda^2_1 \lambda^2_2) + \left( E^2 - v^2_\Delta k^2_\perp \right) \left(\lambda^2_1 + \lambda^2_2 \right)  =0,
\end{eqnarray}
where $\Delta=\sqrt{\mu^2+\Im (\Delta_s)^2}$, and $\tan \phi= \tan \theta^0_{ax}$. By using the Eqs.~(\ref{secularlambda}) and ~(\ref{energylambda}), we obtain the following spectra of the SABS
\begin{equation}\label{surfaceenergy}
E= \pm \sqrt{v^2_\Delta k^2_\perp + \Im (\Delta_s)^2}.
\end{equation}
The explicit form of the $\psi(z)$, as shown in the Supplementary Material \cite{Supplementary}, together with the dispersion $E$ demonstrate that the SABS are massive Majorana fermions, and the gap is given by $\Im (\Delta_s)$. On the other hand, we recover the gapless SABS of the pure B phase~\cite{Tanaka1,Tanaka2}, by setting $\Im (\Delta_s)=0$ in Eq.~(\ref{surfaceenergy}). The gapped SABS have an interesting consequence on the tunneling current measurements. In contrast to the B-phase, there is no ZBCP for the $p+is$ state. Rather, a two gap structure will be found, where the smaller gap stems from the SABS. In the absence of $\mathcal{T}$, the gapless Majorana fermion bound states can only be found along a domain wall between the $p+is$ and the $p-is$ states.

For the characteristic physical response functions of the $p+is$ phase, we first consider the correlation functions of the conserved quantities described by the energy-momentum tensor. In classes D and DIII, the anomalous response of the energy-momentum tensor may be attributed to the gravitational anomaly formula
\begin{equation}
\mathcal{S}_{g}=\frac{1}{1536 \; \pi^2} \int d^4x \; \epsilon^{\alpha \beta \rho \lambda} \; \theta_{ax}(x) \; \mathcal{R}^\eta_{\sigma \alpha \beta} \;  \mathcal{R}^\sigma_{\eta \rho \lambda},
\end{equation}
where $\mathcal{R}_{\eta \sigma \alpha \beta}$ is the Riemann curvature tensor \cite{GravQi,GravRyu1}. Recalling that $\theta^0_{ax}=\pi+\tan^{-1} \Im{(\Delta_s)}/\mu$, we note that the $\pi$ part is tied to the SABS's contribution, whereas $\tan^{-1} \Im{(\Delta_s)}/\mu$ part comes from the scattered states. Recently, it has been argued that the gravitational anomaly may be responsible for a STH effect \cite{Read,GravQi,GravRyu1,Stone}, and some additional cross-correlated responses of the DIII TSC \cite{GravRyu2}, when the $\mathcal{T}$ symmetry is broken on the surface by a weak external Zeeman coupling. The STH conductivity of the massive two dimensional Majorana fermions in the low temperature limit is given by $\kappa_{xy}= \mathrm{sgn}\left(\Im (\Delta_s)\right) \pi T/24$. For the $p+is$ state, the $\mathcal{T}$ is spontaneously broken and no external Zeeman coupling is required to induce this effect. The dimensionless quantities $\hbar \; \kappa_{xy}/(k^2_B T)$ and $\hbar \; \kappa_{xy}/(k_B \mu)$ for the SABS of the $p+is$ state, obtained within a linear response calculation \cite{Murakami} are shown in Fig.~2, as a function $T/T_F$, where $T_F$ is the Fermi temperature. We note that the contribution from the SABS will be generically much larger than that from the scattered states.

Now we briefly discuss the electrodynamic response of this exotic phase. It is natural to anticipate that the signature of the broken $\mathcal{T}$ symmetry can be found through the polar Kerr effect measurements \cite{Kapitulnik}. In addition there will be \emph{dynamic magneto-electric effects}, which can be demonstrated by following the calculations in Refs.~\onlinecite{Witten,GoswamiRoy1}. In these papers, by employing the s-wave and the pseudoscalar pairings of the Dirac fermions, it has been established that the topological electrodynamic response of a TSC is captured by the following magneto-electric term
\begin{equation}
\mathcal{S}_{em}=-\frac{e^2}{64 \pi^2} \int {d^4 x} \; \epsilon^{\mu \nu \rho \lambda} \; \theta_{ax}(x) \; \mathcal{F}_{\mu \nu} \mathcal{F}_{\rho \lambda},\label{eq:14}
\end{equation}
for the massive gauge fields. In the context of $\mathcal{T}$ preserving TSC in class DIII, the observable effect can only come through the surface state contributions (where $\theta_{ax}$ jumps). On the other hand, there are contributions from both the bound and the scattered states for the $p+is$ phase, and $\theta_{ax}(x)$ is dynamical. We also note that in contrast to the \emph{topological magnetic insulators} \cite{Oshikawa}, the massive nature of the gauge field and the axion provides additional stability of this phase. In addition, the gapless one dimensional modes along the line vortex of the B-phase and the pseudoscalar pairings \cite{Silaev, GoswamiRoy2} acquire gap in the $p+is$ state.

We conclude by discussing the experimental prospect of realizing the $p+is$ state. The current experimental status regarding the nature of the paired state in $\mathrm{Cu}_x\mathrm{Bi}_2\mathrm{Se}_3$ is confounding. In Refs.~\onlinecite{PCSasaki, PCKirzhner, PCChen}, a zero bias conductance peak (ZBCP) in point contact spectroscopy measurements has been reported, which is consistent with the existence of the gapless SABS of a TSC. However, the subsequent tunnel spectroscopy measurements on $\mathrm{Cu}_x\mathrm{Bi}_2\mathrm{Se}_3$, with lower copper concentrations, have not found any ZBCP, and the results have been interpreted in terms of the conventional s-wave pairing \cite{ObLevy}. This discrepancy in the spectroscopic measurements on the compounds with different copper concentrations may be an indicator of an underlying competition between the singlet and the triplet pairings. But, the lower quality of the sample currently prohibits a systematic study of the paired state, as a function of the copper concentration. In this direction, the superconducting $\mathrm{Sn}_{1-x}\mathrm{In}_x\mathrm{Te}$ seems to be a promising material, with higher superfluid fraction \cite{ObSasaki}. In Ref.~\onlinecite{ObSasaki} a ZBCP has been reported for $x=0.045$. More recent measurements by Novak {\it et al.}~\cite{Novak} have indicated the existence of a competition between the odd and the even parity pairings in $\mathrm{Sn}_{1-x}\mathrm{In}_x\mathrm{Te}$, which is the crucial ingredient for realizing the $p+is$ phase. In particular, they have argued for a change of the pairing  symmetry around $x=0.038$. We note that the superconductivity in $\mathrm{Sn}_{1-x}\mathrm{In}_x\mathrm{Te}$ is also realized in the ferroelectric phase, which naturally lacks inversion symmetry. The absence of inversion symmetry is conducive for the coexistence of odd and even parity pairings. However it remains to be seen if a $\mathcal{T}$ symmetry broken state is indeed realized in this system, which for example, can be confirmed through surface thermal Hall effect and polar Kerr rotation measurements.

P. G. and B. R. were supported at the National High Magnetic Field Laboratory by NSF Cooperative Agreement No.DMR-0654118, the State of Florida, and the U. S. Department of Energy.

\end{document}